# Text Classification using Artificial Intelligence

S. M. Kamruzzaman

Department of Information and Communication Engineering
University of Rajshahi, Rajshahi-6205, Bangladesh
E-mail: smzaman@gmail.com

## ABSTRACT

**Text classification is the process of classifying documents into predefined categories based on their content. It is the automated assignment of natural language texts to predefined categories. Text classification is the primary requirement of text retrieval systems, which retrieve texts in response to a user query, and text understanding systems, which transform text in some way such as producing summaries, answering questions or extracting data. Existing supervised learning algorithms for classifying text need sufficient documents to learn accurately. This paper presents a new algorithm for text classification using artificial intelligence technique that requires fewer documents for training. Instead of using words, word relation i.e. association rules from these words is used to derive feature set from pre-classified text documents. The concept of naïve Bayes classifier is then used on derived features and finally only a single concept of genetic algorithm has been added for final classification. A system based on the proposed algorithm has been implemented and tested. The experimental results show that the proposed system works as a successful text classifier.**

## 1. INTRODUCTION

There are numerous text documents available in electronic form. More and more are becoming available every day. Such documents represent a massive amount of information that is easily accessible [8]. Seeking value in this huge collection, organization requires much work to organize documents, but this can be automated through data mining-an artificial intelligence technique. The accuracy and understanding of such systems greatly influence their usefulness. The task of data mining is to automatically classify documents into predefined classes based on their content. Many algorithms have been developed to deal with automatic text classification. The most common techniques used for this purpose including naïve Bayes classifier, association rule mining, genetic algorithm, decision tree etc. Association rule mining [1] finds interesting association or correlation among a large set of data items. The discovery of these relationships among huge amounts of transaction records can help in many decision making process. On the other hand, the naïve Bayes classifier uses the maximum a posteriori estimation for learning a classifier. It assumes that the occurrence of each word in a document is conditionally independent of all other words in that document given its class. Although the naïve Bayes works well in many studies [3], [6] it requires a large number of training documents for training accurately. Genetic algorithm starts with an initial population which is created consisting of randomly generated rules. Each rule can be represented by a string of bits. Typically, the fitness of a rule is assessed by its classification accuracy on a set of training examples.

This paper presents a new algorithm for text classification. Instead of using words, word relation i.e. association rules is used to derive feature set from pre-classified text documents. The concept of naïve Bayes classifier is then used on derived features and finally a concept of genetic algorithm has been added for final classification. A system based on the proposed algorithm has been implemented and tested. The experimental results show that the proposed system works as a successful text classifier.

## 2. RELATED WORK

Because of the strong research interest, as shown in the literature, a number of algorithms for text classification have been developed [2]-[3], [5], [9]-[10]. During this work it was tried to consider the established efficient approaches that were found. This section presents a study on some research works to give a view on some techniques used to classify text.

***Bayesian classifiers*** have been gaining popularity lately, and have been found to perform surprisingly well. Classification on new examples is performed with Bayes rule by selecting the class that is most likely to have generated the example. Accuracies found on several research works using naïve Bayes classification were 41% to 74%.

Research on ***text classification using the concept of association rule of data mining*** where naïve Bayes classifier was used to classify text and finally showed the dependability of the naïve Bayes classifier with associated rules [3]. But since this method ignores the negative calculation for any specific class determination in some cases, accuracy may fall. Accuracies found up to 60%.

In ***text classification using decision tree***, authors showed an acceptable accuracy using 76% training data [9], while it is possible to achieve good accuracy using only 40 to 50% of total data as training data.

***Text classification based on genetic algorithm*** showed satisfactory performance using 69% training data, but this process requires time-consuming steps to classify the texts [2], [4].

***Text classification in conference management*** research has emphasized on the effectiveness of supervised machine learning techniques on experimental data sets. In contrast, the amount of research in the area of real-world application is sparse especially in the domain of text classification. In the area of spam detection most of the widely used spam detection products incorporate some kind of personalized text classifier with the majority using incrementally trained naïve Bayes classifiers. In traditional conference management systems problems can occur in the phase where the author has to choose the research topic under which the paper should be filed. The author can be irresolute and uncertain in selecting the topic of the paper. This feeling can be reinforced when the categories are not precisely described. In this system the classifier is trained with examples from previous conferences and so the author should be guided to the correct category. This work shows the average amount of correct classification over all classes reached a value of 74.46% [10].

## 3. BACKGROUND STUDY

### 3.1 Association Rule

Association rule mining finds interesting association or correlation among a large set of data items. In short, association rule is based on associated relationships. Association rules are generated on the basis of two important terms namely minimum support threshold and minimum confidence threshold.

### 3.2 Naïve Bayes Classifier

Bayesian classification is based on Bayes theorem. A simple Bayesian classification namely the naïve classifier is comparable in performance with decision tree and neural network classifiers. Naïve Bayes classifier assumes that the effect of an attribute value on a given class is independent of the values of the other attributes. This assumption is called *class conditional independence*. It is made to simplify the computations involved and, in this sense, is considered "naïve". While applying naïve Bayes classifier to classify text, each word position in a document is defined as an attribute and the value of that attribute to be the word found in that position. Here naïve Bayes classification can be given by:

$V_{NB}$ = argmax P ($V_j$) Π P ($a_j | V_j$).

Here $V_{NB}$ is the classification that maximizes the probability of observing the words that were actually found in the example documents, subject to the usual naïve Bayes independence assumption. The first term can be estimated based on the fraction of each class in the training data. The following equation is used for estimating the second term:

$$\frac{n_k + 1}{n + | vocabulary |} \quad \ldots\ldots\ldots\ldots\ldots\ldots(1)$$

where n is the total number of word positions in all training examples whose target value is $V_j$, $n_k$ is the number of items that word is found among these n word positions, and | vocabulary | is the total number of distinct words found within the training data.

### 3.3 Genetic Algorithm

Genetic algorithms are a part of evolutionary computing, which is a rapidly growing area of artificial intelligence. As we can guess, genetic algorithms are inspired by Darwin's theory of evolution. Simply said, problems are solved by an evolutionary process resulting in a best (fittest) solution (survivor) in other words, the solution is evolved. In general, genetic algorithm starts with an initial population which is created consisting of randomly generated rules. Each rule can be represented by a string of bits. Based on the notion of survival of the fittest, a new population is formed to consist of the fittest rules in the current population, as well as offspring of these rules. Typically, the fitness of a rule is assessed by its classification accuracy on a set of training examples. Offspring are created by applying genetic operators such as crossover and mutation.

## 4. PROPOSED ALGORITHM

The proposed method for classifying text is an implementation of a hybrid method consisting of association rule, naïve Bayes classifier, and genetic algorithm. The features of association rule are used to make association sets. On the other hand, to make a probability chart with prior probabilities, naïve Bayes classifier's probability measurements are used. And finally in the retrieval phase we have implemented the positive-negative matching calculation found in different researches of genetic algorithm [2], [4]. Here the associated word sets, which do not mach with considered class is treated as negative sets and others are positive.

The following algorithm is used for class determination in testing phase.

*n = number of class*
*m = number of associated sets*
   *1. for each class i = 1 to n do*
   *2.   set pval = 0, nval = 0, p = 0, n = 0*
   *3.   for each set s = 1 to m do*
   *4.     if the probability of the class (i) for the set (s) is maximum then increment pval else increment nval*
   *5.     if 50% of the associated set s is matched with the keywords set do step 6 else do step7*
   *6.     if maximum probability matches the class i then increment p*
   *7.     if maximum probability does not match the class i increment n*
   *8.     if (s<=m) go to step 3*
   *9.     calculate the percentage of matching in positive sets for the class i*
   *10.    calculate the percentage of not matching in negative sets for the class i*
   *11.    calculate the total probability as the summation of the results obtained from step 9 and 10 and also the prior probability of the class i in set s*
   *12.    if (i<=n) go to step 1*
   *13.    set the class having the maximum probability value as the result.*

## 5. EXPERIMENTAL EVALUATION

### 5.1 Preparing Text for Classification

Abstracts from different research papers have been used to analyze the experiment. Five classes of papers from physics (PH), chemistry (CH), algorithm (ALG), educational engineering (EDE) and artificial intelligence (AI) were considered for experiment. A total of 296 abstracts (104 from physics, 88 from chemistry, 27 from algorithm, 15 from educational engineering and 62 from AI) were used. The distribution of data is shown in Table 1.

**Table 1:** Distribution of training data.

| Inserted Data to Train | | | | | | |
|---|---|---|---|---|---|---|
| % Data | Total Data | PH | CH | ALG | EDE | AI |
| 15 | 44 | 16 | 13 | 4 | 2 | 9 |
| 20 | 59 | 21 | 18 | 5 | 3 | 12 |
| 25 | 75 | 26 | 22 | 7 | 4 | 16 |
| 30 | 89 | 31 | 26 | 8 | 5 | 19 |
| 35 | 104 | 36 | 31 | 10 | 5 | 22 |
| 40 | 119 | 42 | 35 | 11 | 6 | 25 |
| 45 | 134 | 47 | 40 | 12 | 7 | 28 |
| 50 | 144 | 51 | 43 | 13 | 7 | 30 |
| 55 | 162 | 57 | 48 | 15 | 8 | 34 |

To make the raw text valuable, that is to prepare the text, we have considered only the keywords. That is unnecessary words and symbols are removed. For this keyword extraction process we dropped the common unnecessary words like am, is, are, to, from...etc. and also dropped all kinds of punctuations and stop words. Singular and plural form of a word is considered same. Finally, the remaining frequent words are considered as keywords.

*Let an abstract:* With respect to all algorithm perspective coding binary trees and representation for well-formed parentheses strings. We present here the first Gray code and loopless generating algorithm for P-sequences, and extend them in a Gray code and a new loopless generating algorithm for well-formed parentheses strings. Given a connected graph G = (V,E) and a spanning tree T of G, a fundamental cycle is a cycle resulting by adding an edge e ÎE - T to T. In this paper we establish that the average length of fundamental cycles in a complete graph increases with the number of vertices. Also, given a simple cycle in a complete graph, the paper describes a method of calculating the number of spanning trees, with respect to which the cycle is a fundamental cycle.

Keywords extracted from this abstract are: respect, algorithm, tree, well, formed, parenthese, string, gray, code, loopless, generating, graph, fundamental, cycle, paper, complete, number. This keyword extraction process is applied to all the abstracts.

**Table 2:** Word set with occurrence frequency.

| Large Word Set Found | Number of Occurrence in Documents | | | | |
|---|---|---|---|---|---|
| | PH | CH | ALG | EDE | AI |
| present, well, formed, parentheses, looples | | | 2 | | |
| spanning, tree | | | 2 | | |
| set, length | | | 2 | | |
| educational, significant, study, education, level, student, learning | | | | 3 | |
| handicapped, more, different, environment, effect, working, motivation | | | | 2 | |
| difference, study | | | | 3 | |
| test, significant, difference | | | | 2 | |
| gain, dependencies | | | | | 2 |
| network, neural | | | | | 5 |
| channel, rate | | | | | 2 |
| rate, different | | | | | 3 |
| cold, dark | 2 | | | | |
| obtained, alpha, line | 2 | | | | |
| nuclear, collapse, simulation | 2 | | | | |
| giant, planet, due | 2 | | | | |
| calculate, raman, response, fifth, order | | 2 | | | |
| scalar, included, approximation, dielectric, function | | 2 | | | |
| dirac, fock | | 2 | | | |
| hartree, fock | | 3 | | | |
| excited, using | | 3 | | | |

**Table 3:** Word set with probability value.

| Large Word Set Found | Probability | | | | |
|---|---|---|---|---|---|
| | PH | CH | ALG | EDE | AI |
| present, well, formed, parentheses, looples | 0.013514 | 0.013514 | 0.040541 | 0.013514 | 0.013514 |
| spanning, tree | 0.013514 | 0.013514 | 0.040541 | 0.013514 | 0.013514 |
| set, length | 0.013514 | 0.013514 | 0.040541 | 0.013514 | 0.013514 |
| educational, significant, study, education, level, student, learning | 0.012821 | 0.012821 | 0.012821 | 0.051282 | 0.012821 |
| handicapped, more, different, environment, effect, working, motivation | 0.012821 | 0.012821 | 0.012821 | 0.038462 | 0.012821 |
| difference, study | 0.012821 | 0.012821 | 0.012821 | 0.051282 | 0.012821 |
| test, significant, difference | 0.012821 | 0.012821 | 0.012821 | 0.038462 | 0.012821 |
| gain, dependencies | 0.012346 | 0.012346 | 0.012346 | 0.012346 | 0.037037 |
| network, neural | 0.012346 | 0.012346 | 0.012346 | 0.012346 | 0.074074 |
| channel, rate | 0.012346 | 0.012346 | 0.012346 | 0.012346 | 0.037037 |
| rate, different | 0.012346 | 0.012346 | 0.012346 | 0.012346 | 0.049383 |
| cold, dark | 0.034483 | 0.011494 | 0.011494 | 0.011494 | 0.011494 |
| obtained, alpha, line | 0.034483 | 0.011494 | 0.011494 | 0.011494 | 0.011494 |
| nuclear, collapse, simulation | 0.034483 | 0.011494 | 0.011494 | 0.011494 | 0.011494 |
| giant, planet, due | 0.034483 | 0.011494 | 0.011494 | 0.011494 | 0.011494 |
| calculate, raman, response, fifth, order | 0.010638 | 0.031915 | 0.010638 | 0.010638 | 0.010638 |
| scalar, included, approximation, dielectric, function | 0.010638 | 0.031915 | 0.010638 | 0.010638 | 0.010638 |
| dirac, fock | 0.010638 | 0.031915 | 0.010638 | 0.010638 | 0.010638 |
| hartree, fock | 0.010638 | 0.031915 | 0.010638 | 0.010638 | 0.010638 |
| excited, using | 0.010638 | 0.042553 | 0.010638 | 0.010638 | 0.010638 |

**5.2 Training Phase**

**5.2.1 Deriving associated word sets**

Each abstract is considered as a transaction in the text data. After pre-processing the text data association rule mining [1] is applied to the set of transaction data where each frequent word set from each abstract is considered as a single transaction. Using these transactions, we generated a list of maximum length sets applying the apriori algorithm [1]. The support and confidence is set to 0.05 and 0.75 respectively. A list of the generated large word set for 55% of training data with their occurrence frequency is illustrated in Table 2.

**5.2.2 Setting associated word set with probability value**

To use the naïve Bayes classifier for probability calculation the generated associated sets are required. The calculation of first term of this classifier is based on the fraction of each target class in the training data. From the generated word set after applying association mining on training data we have found the following information:

Total number of word set = 69

Total number of word set from Physics (PH) = 17, Chemistry (CH) = 25, Algorithm (ALG) = 5, Educational Engineering (EDE) = 9, and Artificial Intelligence (AI) = 12.

Prior probability we had for PH, CH, ALG, EDE and AI are 0.26, 0.36, 0.07, 0.13 and 0.17 respectively. Then the second term is calculated according to the equation (1). The probability values of word set are listed in Table 3.

### 5.3 Testing Phase

#### 5.3.1 Classifying a new document

*Let an abstract:* The dielectric function of heavy nonmetallic crystals are studied within a relativistic framework using the ADF-BAND program package. The calculations are based on the work that has been done to calculate the dielectric response of nonmetallic crystals. The starting point of the relativistic corrections is the Dirac equation in an quasi-static electric field. As the Dirac equation is a four-component equation it is first reduced to a two-component equation with the Foldy-Wouthuysen transformation. The then obtained two-component Dirac-Hamiltonian is then used to find (after some treatments of this Hamiltonian) an expression for the matrixelements required With these matrixelements the dielectric function can be evaluated, but now relativistically corrected. The obtained relativistic corrected dielectric function was finally evaluated for some light crystals; C, Si, Ga, As and He and for heavier crystals asto see if the relativistic corrections indeed improve on the dielectric function of the studied crystals. The heavy crystals with large errors as compared to experiment were studied.

The extracted sets of keywords of a new abstract are: dielectric, function, heavy, nonmetallic, crystal, studied, relativistic, article, correction, dirac, equation, component, two, obtained, hamiltonian, some, matrix, element, evaluated, corrected.

All these keywords are sent to the classifier where for each class a common circle runs. As a result of this run a probability is obtained for each class. As for example for the class Chemistry the given set gives: pval=25, nval=44, p=2, n=43. Now the probability of CH = ((p*100)/pval) + ((n*100)/nval) + prior probability of CH
= ((2*100)/25) + ((43*100)/44) + 0.36 = 106.09

For this set of keywords, Calculated Probability for class PH=101.89, CH=106.09, ALG=95.38, EDE=95.13 and AI=94.91. From the above result we found the abstract belongs to class Chemistry (CH).

**Table 4:** Accuracy Vs training data.

| % of Training Data | % of Accuracy |
|---|---|
| 10 | 31 |
| 15 | 34 |
| 20 | 32 |
| 25 | 54 |
| 30 | 57 |
| 35 | 70 |
| 40 | 68 |
| 45 | 68 |
| 50 | 80 |
| 55 | 68 |

Table 4 shows the percentage of accuracy Vs percentage of training data. At the beginning of the experiment, we started with 10% of the total data to train, which showed unsatisfactory accuracy. Then we increased training data to 20%, which showed development in accuracy. Next as we increased the percentage of training data accuracy became more desirable. We checked up to 55% training data. In this process, considering overall accuracy we choose 79% accuracy i.e., 50% training data as the best.

### 6. COMPARATIVE STUDY

The comparisons of the proposed method to the related works are given below:

**Naïve Bayes text classification**
The reported results shown in [7] that the multinomial event model reaches a maximum of 54% accuracy at a vocabulary size of 1000 words. The multivariate Bernoulli event model reaches a maximum of 41% accuracy with only 200 words. The multinomial has the highest accuracy of 74% at 2000 words, and multivariate Bernoulli is best with 46% accuracy at 1000 words. Here in all cases 70% of the data used for training.

**Association rule and naïve Bayes classifier**
The following results are found using the same data sets for both association rule with naïve Bayes classifier [3] and proposed method. It is seen in Fig. 1 that proposed approach work well using only 50% training data.

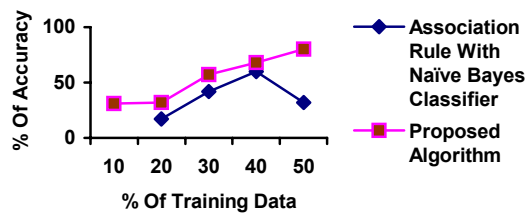

**Fig. 1:** Accuracy Vs training curve.

**Association rule based decision tree**

In text categorization using association rule based decision tree [9], 76 % data set of the total 33 data set were used to train and 13% error observed. On the other hand using only 50% data as training the proposed algorithm can able to classify text with 80% accuracy rate. The major problem of decision tree based classified [9] is that, this system is totally failed to categorize a class. Our proposed technique shows better performance even with 8 times larger data sets.

**Text classification using genetic algorithm**

Researchers showed 68% accuracy using the concept of genetic algorithm with 31% testing data [4] while our technique is better shows 80% accuracy with 50% of testing data. Moreover text classification using genetic algorithm require processing for each class during training. But our proposed algorithm does not require such process during training phase and hence reduces time.

**Text classification in conference management**

This paper presented an approach to integrate an automatic text classification system into a conference management system [10]. Results showed that in categories where enough good training examples were present the user did not change the automatically pre-selected category that often. Another implication is that classes that overlap with other classes or are subclasses of others perform quite poorly. In general 2/3 of the classification results were not changed by the users in the first scenario. In the second scenario, where more documents per class were available, the average amount of correct classification over all classes reached a value of 74.46%. Whereas our proposed system shows accuracy 80%, which is far better than this system.

**Table 5:** Comparison of proposed method with other methods.

| Technique | (%) Training Data | (%) Highest Accuracy |
|---|---|---|
| Naïve Bayes | 70 | 74 |
| Association rule and Naïve Bayes | 50 | 60 |
| Association Rule Based Decision Tree | 76 | 87 |
| Genetic Algorithm | 69 | 68 |
| Conference Management | 70 | 76.46 |
| **Proposed Algorithm** | **50** | **80** |

It is shown in Table 5 that the proposed algorithm outperforms all the other algorithms with a marginal difference except association rule based Decision Tree where 76% data was used for training purpose.

## 7. CONCLUSION

This paper presented an efficient technique for text classification. The existing techniques require more data for training as well as the computational time of these techniques is also high. In contrast to the existing algorithms, the proposed hybrid algorithm requires less training data and less computational time. In spite of the randomly chosen training set we achieved 90% accuracy for 50% training data. Though the experimental results are quite encouraging, it would be better if we work with larger data sets with more classes.